\pdfoutput=1

\documentclass[10pt,conference,letter]{IEEEtran}
\usepackage{blindtext, graphicx}
\usepackage{times}

\IEEEoverridecommandlockouts
\usepackage{cite}
\usepackage{amsmath,amssymb,amsfonts}
\usepackage{algorithmic}
\usepackage{textcomp}
\usepackage{setspace}
\def\smallerspacecaption{\vspace{-2mm}}
\def\smallspaceenum{\vspace{1.25mm}}

\usepackage{floatrow}
\usepackage[caption=false,font=footnotesize]{subfig}

\usepackage[table, dvipsnames]{xcolor}  
\usepackage{fancyhdr}
\usepackage{graphicx}
\usepackage{url}
\usepackage{verbatim}
\usepackage{multirow}
\usepackage{hhline}
\usepackage[us]{datetime}
\usepackage{paralist}
\usepackage{color}
\usepackage{soul}
\usepackage[implicit=false,hidelinks]{hyperref}
\usepackage{stfloats}
\usepackage[ruled, vlined, norelsize]{algorithm2e} 
\SetAlFnt{\sf} 

\graphicspath{{incl/}}

\usepackage{soul}
\newcommand{\drop}[1]{\textcolor{red}{#1}}
\renewcommand{\drop}[1]{}

\newdimen\arrayruleHwidth
\makeatletter
\def\Hline{\noalign{\ifnum0=`}\fi\hrule \@height \arrayruleHwidth
\futurelet \@tempa\@xhline}
\makeatother

\makeatletter
\def\blfootnote{\xdef\@thefnmark{}\@footnotetext}
\makeatother

\shortdate
\settimeformat{ampmtime}

\makeatletter
\newcommand*\tabsize{%
	   \@setfontsize\tabsize{6}{7.2}%
}
\makeatother

\renewcommand{\arraystretch}{1.03}

\renewcommand{\IEEEbibitemsep}{0.3pt}
\makeatletter
\IEEEtriggercmd{\reset@font\normalfont\fontsize{6.8pt}{7.11pt}\selectfont}
\makeatother
\IEEEtriggeratref{1}

\begin{document}
\title{
Advancing Hardware Security Using Polymorphic and Stochastic Spin-Hall Effect Devices
}

\newcommand*\samethanks[1][\value{footnote}]{\footnotemark[#1]}

\author{
	{ Satwik Patnaik$^{\dagger}$\textsuperscript{*}\thanks{\textsuperscript{*}S.\ Patnaik and N.\ Rangarajan contributed equally.}, Nikhil Rangarajan$^{\dagger}$\textsuperscript{*}, Johann Knechtel$^{\ddagger}$,
	   Ozgur Sinanoglu$^{\ddagger}$,}{ and Shaloo Rakheja$^{\dagger}$}\\[2pt]
  {$^{\dagger}$\,Tandon School of Engineering, New York University, New York, USA}\\
  {$^{\ddagger}$\,Division of Engineering, New York University Abu Dhabi, Abu Dhabi, United Arab Emirates}\\
  \normalsize{\{sp4012, nikhil.rangarajan, johann, ozgursin, shaloo.rakheja\}@nyu.edu}
}

\maketitle

\renewcommand{\headrulewidth}{0.0pt}
\thispagestyle{fancy}
\pagestyle{fancy}
\cfoot{
	\vspace{-1cm}
\copyright~2018 IEEE.
This is the author's version of the work. It is posted here for your personal use.
	Not for redistribution.\\
	The definitive Version of Record is published in
	Proc. Design, Automation and Test in Europe (DATE) 2018\\
}

\begin{abstract}
Protecting intellectual property (IP) in electronic circuits has become a serious challenge in recent years.
Logic locking/encryption and layout camouflaging are two prominent techniques
for IP protection.
Most existing approaches, however, particularly those focused on CMOS integration, incur excessive design overheads resulting from their need for additional circuit structures or device-level modifications.
This work leverages the innate polymorphism of an emerging spin-based device, called the giant spin-Hall effect (GSHE) switch, to simultaneously enable locking and camouflaging within a single instance.
Using the GSHE switch, we propose a
powerful primitive that enables cloaking all the 16 Boolean functions possible for two inputs.
We conduct a comprehensive
study using state-of-the-art Boolean satisfiability (SAT) attacks to
demonstrate the superior resilience of the proposed primitive in comparison to several others in the literature.
While we tailor the primitive for deterministic computation, it
can readily support stochastic computation; we argue that stochastic behavior
can break most, if not all, existing SAT attacks.
Finally, we discuss the resilience of the primitive against various side-channel attacks as well as
invasive monitoring
at runtime, 
which are arguably even more concerning threats than SAT attacks.

\end{abstract}

\section{Introduction}
\label{introduction}
With the advent of globalization affecting the design and manufacturing process of integrated circuits (ICs),
hardware security has emerged as a critical concern.
The exposure to various adversaries, which may reverse engineer (RE) ICs, counterfeit them, steal their intellectual property (IP), inject hardware Trojans, leak and/or extract
sensitive data at runtime has escalated~\cite{rostami14}.
Next, we briefly review IP protection schemes and attacks in general.

\textbf{IC camouflaging} seeks
to mitigate RE attacks, wherein the layout-level appearance of the IC is altered such that it becomes intractable to
decipher its underlying functionality and IP.
For CMOS integration, various techniques have been proposed, e.g., look-alike gates~\cite{rajendran13_camouflage},
threshold-dependent camouflaging~\cite{nirmala16, erbagci16}, and obfuscated interconnects~\cite{patnaik17_Camo_BEOL_ICCAD}.

\textbf{Logic locking/encryption} 
obfuscates the IP functionality rather than the device-level 
layout~\cite{yasin16_SARLock,
		xie16_SAT}.
The so-called key gates are carefully tailored into the IP/chip,
where only the correct key can ``unlock'' the original functionality.

\textbf{Analytical attacks} targeting camouflaged (or locked) ICs were initially introduced in~\cite{subramanyan15, massad15}.
These attacks are based on Boolean satisfiability (SAT) and the fact that a small set of discriminating input patterns (DIPs)
may suffice to infer the camouflaged functionality (or locking key).
Several
SAT-attack resilient techniques were recently proposed~\cite{yasin16_SARLock, xie16_SAT,
	li16_camouflaging}; however, most of these techniques are still
vulnerable to advanced analytical attacks such as~\cite{shamsi17, shen17, bypass-attack2017}.

\textbf{Physical attacks} range from non-invasive (e.g., power side-channel attacks) and semi-invasive (e.g., localized fault-injection attacks) to invasive attacks (e.g., RE,
		microprobing the frontside/backside)~\cite{Wang17probing}.
Such attacks are also promising for extracting sensitive data at runtime, even from secured chips, e.g.,~\cite{skorobogatov12,courbon16}.

\textbf{Emerging devices} including, e.g.,
nanowire transistors,
carbon-based or spin-based devices,
	may offer
lower power dissipation and higher integration density compared to their CMOS counterparts~\cite{nikonov2013overview}.
Additionally,
   emerging devices can augment the CMOS technology to improve
   hardware security~\cite{ghosh2016spintronics,bi16_JETC, parveen2017hybrid}.
The most promising aspect of 
many
emerging devices is \emph{polymorphism}: a polymorphic gate can readily implement different Boolean functions at runtime, where the functionality is determined by an
internal/external control mechanism~\cite{
		parveen2017hybrid}.
It is important to note that polymorphic gates
can inherently support both
camouflaging and locking due to the following reasons.
First, owing to their uniform device-level layout, the actual function of a polymorphic gate is hard to determine,
	particularly when optical-imaging-based RE techniques are used.
	Second, the actual function
	is dependent on the control input, which can act as a key input.

\textbf{In this work}, we 
use the giant spin-Hall effect (GSHE) switch, 
first proposed in~\cite{datta2012non}, to build polymorphic gates
for advanced protection.
More specifically, we leverage the GSHE switch
recently designed and analyzed
by Rangarajan \emph{et al.\ }\cite{rangarajan2017energy}
in the context of probabilistic computing.
We emphasize that the notions of locking and camouflaging are
interchangeable in this work due to the polymorphic nature of the proposed primitive,
unlike for CMOS-centric approaches.
The contributions of this work can be summarized as follows.
\begin{enumerate}
\item We leverage a polymorphic, GSHE-based device to propose
a versatile security primitive.
The primitive provides strong camouflaging capabilities---given two inputs, all 16 possible Boolean functions can be cloaked within a single instance.
We elaborate on the device as well as the proposed primitive in detail in Sec.~\ref{device_model}.
\smallspaceenum
\item We analyze the protection provided by the primitive 
against
attacks such as imaging- and electron-microscopy-based RE,
side-channel attacks,
and analytical SAT attacks (Sec.~\ref{security}).
As for SAT attacks, a comprehensive study is conducted and benchmarked against prior state-of-the-art techniques.
Immunity to SAT attacks for probabilistic computing, directly supported by the primitive, is also discussed.
\smallspaceenum
\item We outline the prospects of
	hybrid CMOS-GSHE designs for
		industrial benchmarks.
We observe that delay-aware protection can provide strong resilience (against SAT attacks) with negligible layout overheads.
\end{enumerate}

\section{Background: Prior Art and Limitations}
\label{sec:background}

In~\cite{zhang2015giant}, the authors 
implemented a low-power and versatile gate
using a GSHE-based magnetic tunnel junction (MTJ) as the basic switching element.
However, this device is not explicitly tailored for security;
it is unable to support logic locking by itself, as it is
not truly polymorphic.
More concerning
is the limitation to only four possible Boolean functions, which renders this primitive weak against SAT attacks (Sec.~\ref{security}).

Alasad~\emph{et al.}~\cite{alasad2017leveraging} use all-spin logic (ASL) to design 
three different security primitives, supporting three sets of camouflaged functionalities: INV/BUF, XOR/XNOR, and AND/NAND/OR/NOR.
The layouts of the three primitives are unique; they can be readily distinguished
by imaging-based RE tools, which also eases subsequent SAT attacks
(Sec.~\ref{security}).

Winograd~\emph{et al.}~\cite{winograd2016hybrid} introduced a spin-transfer torque (STT)-based reconfigurable lookup table (LUT), explicitly addressing hardware security.
However, their approach
	falls short in terms of resilience against SAT attack.
(Note that the authors did not report on any SAT attack themselves.)
	We protect the \emph{s38584} benchmark
	according to their technique
and observe that the protected layout can be decamouflaged in less than 30 seconds on average (over 100 runs of camouflaging and SAT attacks).
This weak resilience stems from the limited use of their STT-LUT primitive
to curb power, performance, and area (PPA) overheads.

As for CMOS-centric camouflaging,
most schemes
incur a high layout cost. For example,
the look-alike NAND-NOR-XOR gate proposed by Rajendran~\emph{et al.}~\cite{rajendran13_camouflage} induces 4$\times$ area, 5.5$\times$ power, and 1.6$\times$ delay (compared to a
		regular two-input NAND gate)
whereas the threshold-dependent full-chip camouflaging as proposed in \cite{erbagci16} still induces overheads of 14\%, 82\%, and 150\% in PPA, respectively.
As a result, most schemes are limited to a cost-constrained and selective application, which has severe implications for security
(Sec.~\ref{security}).

\section{Device-Level Design of Spin-Based Primitive}
\label{device_model}

Protection schemes based on emerging devices can be competitive, even when compared to regular CMOS.
While the GSHE switch leveraged in this work is still in the nascent stage of fabrication~\cite{penumatcha2016impact},
it is nevertheless promising because of its small scale and low power 
(Section~\ref{properties_GSHEswitch}).
As for the relatively large delay,
the GSHE-based primitive is still applicable without inducing significant delay overheads
(Sec.~\ref{security}).

\subsection{Structure and Operating Principle of the GSHE Switch}
\begin{figure}[tb]
    \centering
    \includegraphics[width=.89\textwidth]{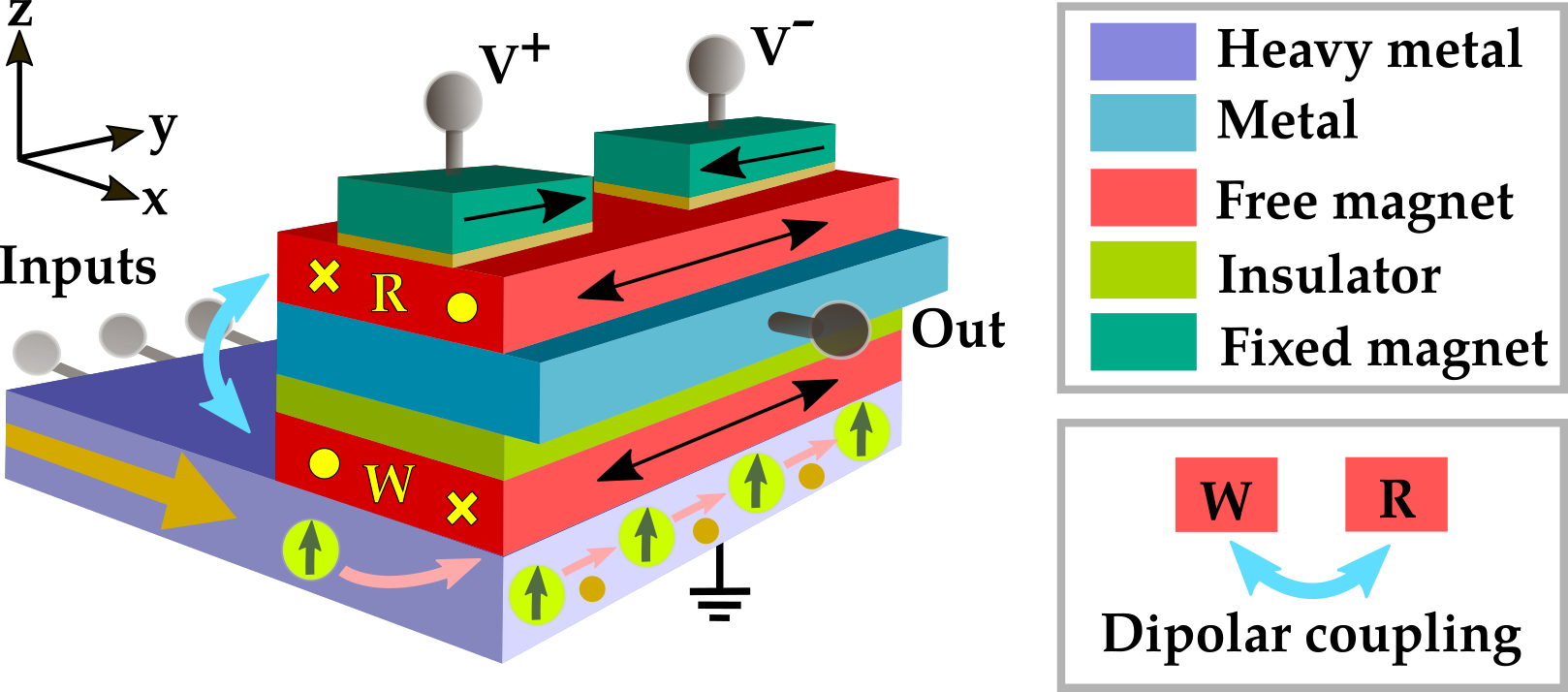}
    \caption{
Structure of the GSHE switch. The concept is derived from~\cite{rangarajan2017energy}, but here we adopt a stacked integration to maximize the dipolar coupling.}
    \label{GSHE_switch}
    \smallerspacecaption
\end{figure}

The GSHE switch, which is at the heart of the proposed primitive, is shown in Fig.~\ref{GSHE_switch}.
Above the heavy metal spin-Hall layer (purple, bottom) are the write (W; red, bottom) and read (R; red, top) nanomagnets (NM).
These nanomagnets (W-NM and R-NM) exhibit a negative mutual dipolar coupling.
On top of the R-NM sit two fixed ferromagnetic layers (dark green) with anti-parallel magnetization
directions.

\begin{figure}[tb]
    \centering
    \includegraphics[width=\textwidth]{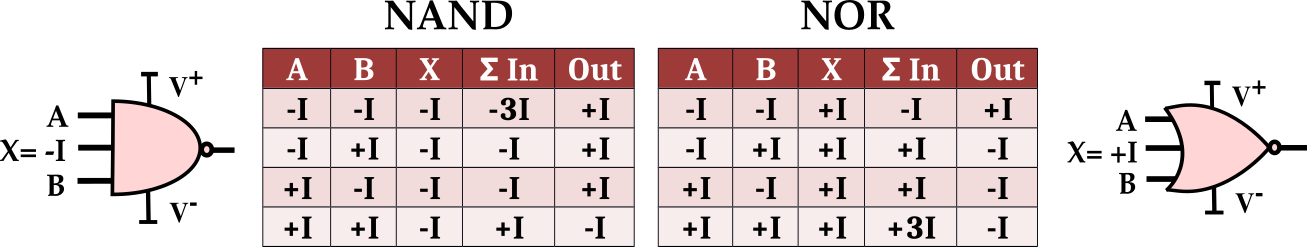}
    \caption{The current-centric truth tables for NAND and NOR functionalities, with inputs A and B (X is a control signal). As always the case for our GSHE-based primitive, logic 1/0 is represented by an output current +I/-I.
    \label{GSHE_NAND_NOR}
    }
    \smallerspacecaption
\end{figure}

Applying a charge current to the bottom layer (large golden arrow in Fig.~\ref{GSHE_switch}) results in spin accumulation of one polarity (green spin-up spheres)
in the transverse direction (pink arrows)~\cite{rangarajan2017energy}.
This spin-polarized current then imparts a spin-transfer torque (STT) to the W-NM~\cite{slonczewski1996current}. The STT switches the W-NM from one stable state to the other which, in turn,
switches the R-NM in the opposite sense.\footnote{That is because in the presence of negative magnetic
		dipolar coupling, the minimum energy state is the one in which the W and R
	nanomagnets are anti-parallel to each other~\cite{datta2012non}.
\vspace{-1ex}
}
Now, the magnetization direction of the R-NM
will be parallel to one of the fixed ferromagnets on top and anti-parallel to the other.
The parallel path offers a lower resistance for a charge current passing from/to the respective top contact to/from the output terminal.
This read-out phase
commences once voltages are applied to the top contacts ($V^+$ and $V^-$).
Depending on the polarity of the voltage applied to the low-resistance path, the output current either flows inward or outward---this represents the binary result of the GSHE
switch operation (see Fig.~\ref{GSHE_NAND_NOR}).

\subsection{Characterization and Comparison of the GSHE Switch}
\label{properties_GSHEswitch}
\begin{figure}[h]
\centering
\includegraphics[width=.75\textwidth]{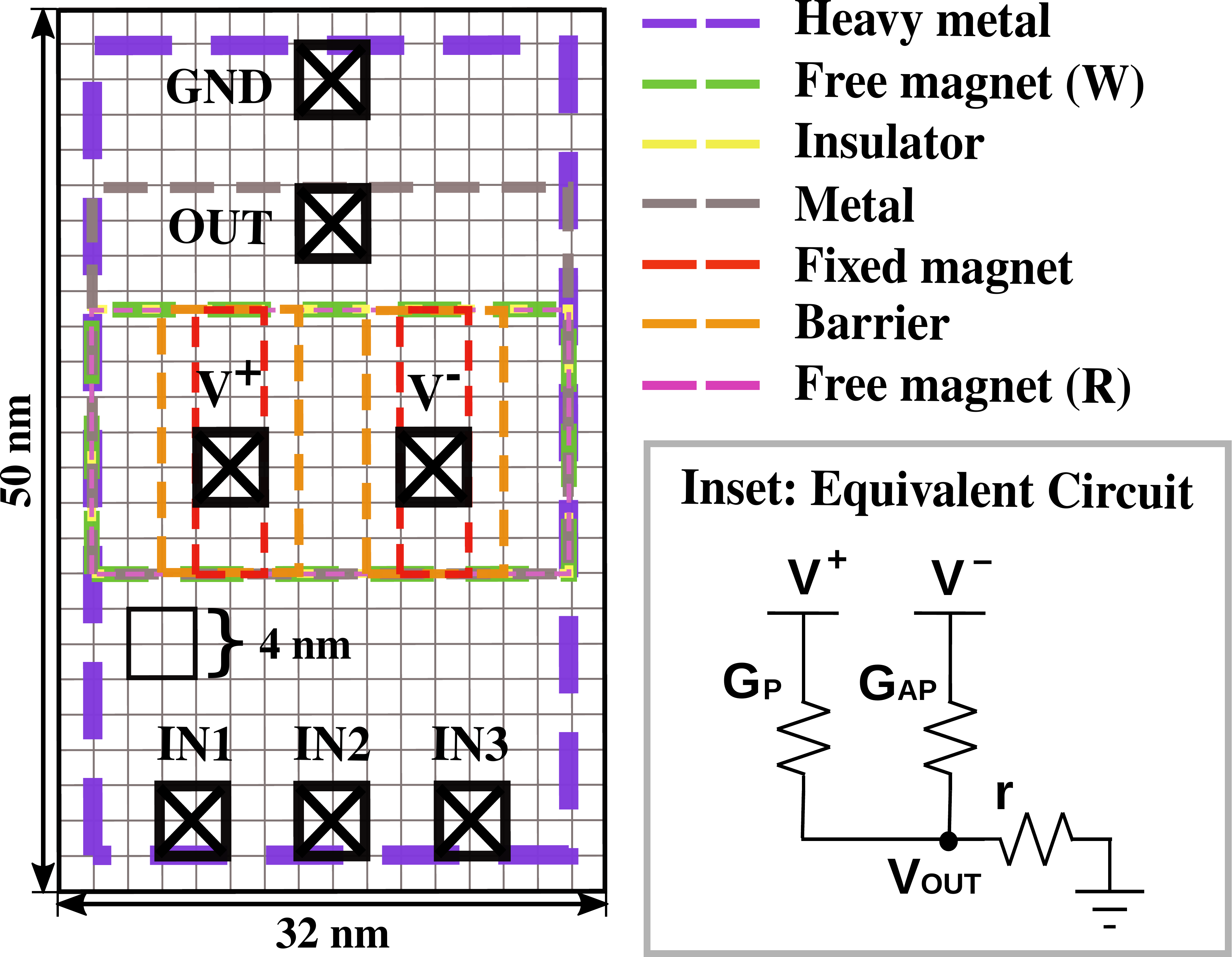}
\smallerspacecaption
\caption{The conceptual layout of the GSHE switch (main part), and the equivalent circuit (inset, derived from~\cite{datta2012non}).
	The power dissipation of the latter
		is dictated by
the resistance $r$ of the heavy metal as well as the conductances of the anti-parallel, high-resistance path ($G_{AP}$) and the parallel, low-resistance path ($G_P$) build up by the fixed
ferromagnets.}
\label{fig:Layout_GSHE}
\end{figure}

The conceptual layout of the GSHE switch 
(Fig.~\ref{fig:Layout_GSHE}) is drawn based on the design rules for beyond-CMOS devices~\cite{nikonov2013overview}, i.e., in units of maximum misalignment length
$\lambda$. The
area of the GSHE switch is accordingly estimated to be $0.0016 \mu$m$^2$.
The material parameters are given in Table~\ref{parameters}.
Notably, 
a
spin current ($I_{S}$) of at least 20$\mu$A is required in this work to guarantee a deterministic switching behavior.

\begin{table}[tb]
\scriptsize
\renewcommand{\arraystretch}{1.1}
\caption{Material Parameters of the GSHE Switch
}
\smallerspacecaption
\begin{center}
\setlength{\tabcolsep}{1.4mm}
\begin{tabular}{c|c}
\hline
{\textbf{Parameter}} & {\textbf{Value}}  \\
\hline
\hline
Volume of nanomagnets (NM) & ($28\times 15\times 2$) nm$^3$~\cite{rangarajan2017energy} \\
\hline
\multirow{2}{*}{Saturation magnetization $M_s$ of NM } &  $10^6$ A/m (W-NM)~\cite{rangarajan2017energy}\\
	   & $5\times 10^5$ A/m (R-NM)~\cite{rangarajan2017energy}\\
\hline
\multirow{2}{*}{Uniaxial energy density $K_u$ of NM} & $2.5\times 10^4$ J/m$^3$ (W-NM)~\cite{rangarajan2017energy} \\
& $5\times 10^3$ J/m$^3$ (R-NM)~\cite{rangarajan2017energy} \\
 \hline
Spin current $I_{S}$, determ.\ switching
& 20 $\mu$A~\cite{rangarajan2017energy}\\
\hline
Resistance area product $RAP$  &$1\>\> \Omega \mu$m$^2$ \cite{maehara2011tunnel}\\
\hline
Tunneling magnetoresistance $TMR$ &$170\%$ \cite{maehara2011tunnel}\\
\hline
Parallel conductance $G_P$ & $420\>\>\mu$S\\
\hline
Anti-parallel conductance $G_{AP}$ & $155.6\>\>\mu$S\\
\hline
Resistivity of heavy metal (HM) $\rho$ &$5.6\times 10^{-7} \Omega$--m\\
\hline
Spin-Hall angle $\theta_{SH}$ of HM &$0.4$\\
\hline
Thickness $t_{HM}$ of HM & $1$ nm\\
\hline
Internal gain $\beta$ of HM & $0.4\times(15\;$nm$/1\;$nm$)$\\
$\beta = \theta_{SH}\times(w_{NM}/t_{HM})$ & $=6$ \\
\hline
Resistance $r$ of HM & $\approx 1\>\> k\Omega$ \\
\hline
\end{tabular}
\label{parameters}
\end{center}
\end{table}

\begin{figure}[tb]
\centering
\includegraphics[width=.85\textwidth]{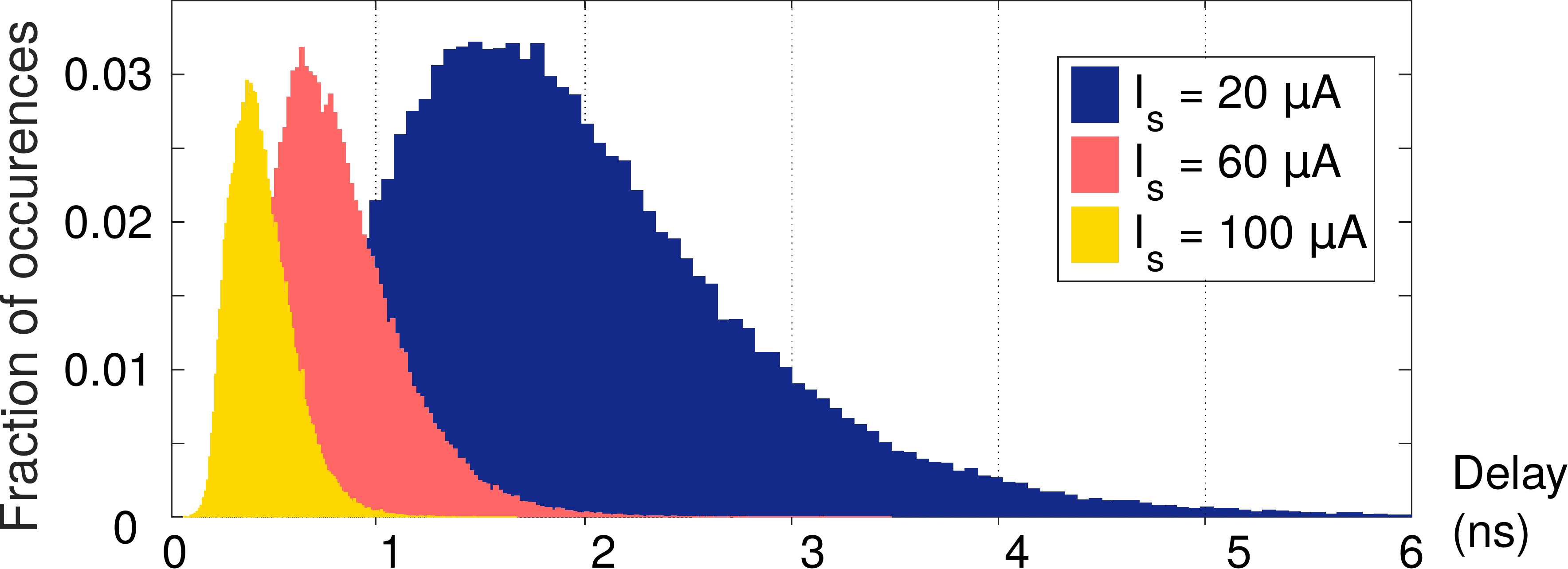}
\smallerspacecaption
\caption{Delay distributions for the GSHE switch at various spin currents ($I_{S}$).
The distributions are obtained from 100,000 simulations. 
Although the delays incurred in switching are stochastic, the switching process itself is still deterministic.
Note that the spread and mean delay diminish with increasing $I_{S}$, however, at the cost of
higher power dissipation.}
\label{fig:delay_profile}
\end{figure}

The performance of the switch is determined by the nanomagnetic dynamics, which is simulated using the stochastic Landau-Lifshitz-Gilbert-Slonczewski equation~\cite{d2006midpoint}. 
Three simulated delay distributions
are illustrated in Fig.~\ref{fig:delay_profile}.
For the propagation delay of the primitive,
   we subsequently assume a mean delay of
1.55~ns obtained for $I_{S} = 20$~$\mu$A.

The power dissipation
for the read-out phase is
derived according to the equivalent circuit shown in Fig.~\ref{fig:Layout_GSHE} (inset).
Using the following equations and the parameters listed in Table~\ref{parameters}, the power dissipation of the GSHE switch (including leakage) is derived as 0.2125 $\mu$W.

\vspace{-2ex}
\begin{subequations}
\small
\begin{equation*}
P = \frac{{V_{OUT}}^{2}}{r} + (V_{SUP}-V_{OUT})^{2}G_{P} + (V_{OUT}+V_{SUP})^{2}G_{AP}
\end{equation*}
\begin{equation*}
V_{SUP} = \left|V^{+/-}\right| = \left(\frac{I_{S}}{\beta}\right)\left(\frac{1+r(G_{P}+G_{AP})}{G_{P}-G_{AP}}\right); \; V_{OUT} = \frac{I_{S}\>\> r}{\beta}
\end{equation*}
\begin{equation*}
\frac{G_{P}}{G_{AP}} = 1 + TMR; \; G_{P} = \frac{A(nanomagnets)}{RAP}
\end{equation*}
\end{subequations}

In Table~\ref{tab:devices}, we compare the metrics of the GSHE switch against those of existing devices, including ones that are not necessarily security-oriented.
The switch is superior in terms of energy/power
	but is limited in terms of delay.
	As for security, the number of possible functions is the
relevant metric; here, the GSHE switch significantly outperforms prior art. 
Moreover, a delay-aware application can provide adequate security without any significant overheads (Sec.~\ref{security}).

\begin{table}[tb]
\centering
\scriptsize
\caption{Comparison of Selected Emerging-Device Primitives}
\label{tab:devices}
\smallerspacecaption
\setlength{\tabcolsep}{1.2mm}
\begin{tabular}{c|c|c|c|c}
\hline
\textbf{Publication} & \textbf{\# Functions} & \textbf{Energy} & \textbf{Power} & \textbf{Delay}\\
\hline 
\hline
~\cite{bi16_JETC} SiNW & NAND/NOR & 0.05--0.1 fJ & 1.13--1.77 $\mu$W & 42--56 ps \\ \hline
~\cite[a]{alasad2017leveraging} ASL & NAND/NOR/AND/OR & 0.58 pJ & 351.52 $\mu$W & 1.65 ns \\ \hline
~\cite[b]{alasad2017leveraging} ASL & XOR/XNOR & 1.16 pJ & 351.52 $\mu$W  & 3.3 ns \\ \hline
~\cite[c]{alasad2017leveraging} ASL & INV/BUF & 0.13 pJ & 342.11 $\mu$W & 0.38 ns \\ \hline
~\cite{huang2016magnetic} DWM  & AND/OR & 67.72 fJ & 60.46 $\mu$W & 1.12 ns \\ \hline
\multirow{2}{*}{~\cite{parveen2017hybrid} DWM}  &  NAND/NOR/XOR/XNOR/ & \multirow{2}{*}{N/A} & \multirow{2}{*}{N/A} & \multirow{2}{*}{N/A}	\\
&  AND/OR/INV
	& & & \\ \hline
~\cite{zhang2015giant} GSHE  & 	AND/OR/NAND/NOR & N/A	& N/A		& N/A	\\ \hline
\multirow{2}{*}{~\cite{winograd2016hybrid} STT} &  NAND/NOR/XOR/ & \multirow{2}{*}{N/A}	& \multirow{2}{*}{N/A}		& \multirow{2}{*}{N/A}	\\
&  XNOR/AND/OR & & & \\ \hline
\textbf{This work} & \textbf{All 16} & \textbf{0.33 fJ} & \textbf{0.2125 $\mu$W} & \textbf{1.55 ns} \\ \hline

\end{tabular}
\end{table}

\subsection{Security Primitive: Cloaking of all 16 Boolean Functions}
\begin{figure*}[tb]
    \centering
    \includegraphics[width=.9\textwidth]{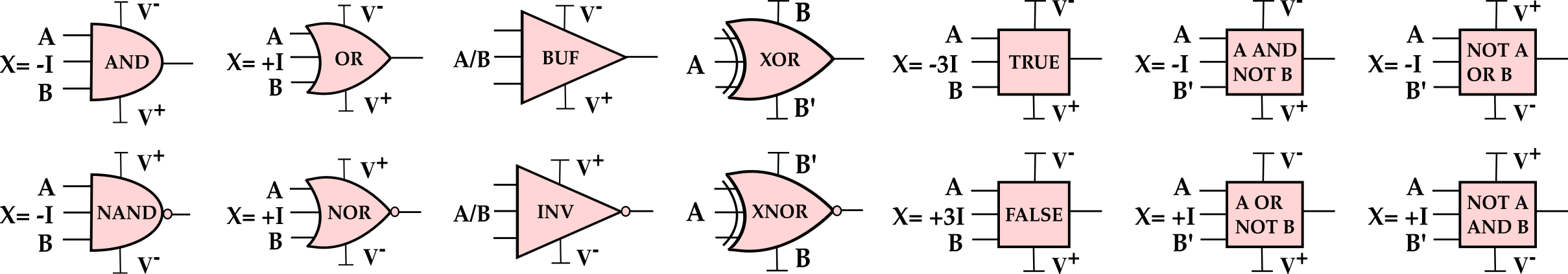}
    \caption{All 16 possible Boolean functionalities for two inputs, A and B, implemented using the proposed primitive.
	    If required, X serves as control signal, not as regular input. Note that BUF and INV capture two functionalities each.
    \label{fig:GSHE_gates}
    }
    \smallerspacecaption
\end{figure*}

All 16 possible Boolean functions implemented by the proposed primitive are illustrated in Fig.~\ref{fig:GSHE_gates}.
To realize NAND/NOR, e.g., three charge currents are fed into the bottom layer of the GSHE switch at once:
two currents represent the logic signals A and B, and the third current (X) acts as the ``tie-breaking'' control input
(recall Fig.~\ref{GSHE_NAND_NOR}).
For the XOR/XNOR functionalities,
one signal is provided as input current, whereas the other signal and its inverse are provided as input voltages at the $V^+$ and $V^-$ terminals of the fixed
ferromagnets.\footnote{Toward this end, magneto-electric
		transducers~\cite{manipatruni15} may be placed in the interconnects.
		Such transducers can be tailored for uniform, indistinguishable layouts, and can be used to convert (i)~charge currents to their reverse (+I to -I, or B to B'), (ii)~voltages
		to charge currents (high/low voltage to +/-I),
				and (iii)~charge currents to voltages (+/-I to high/low voltages).
}
Swapping the voltage polarities
switches between the complementary functions.

Note that 
three wires are used for the input terminal for all 16 Boolean gates (recall Fig.~\ref{fig:Layout_GSHE}); 
this
renders the layout of the primitive indistinguishable for optical-imaging-based RE, irrespective of the actual functionality. As such, some gates will 
require dummy wires.
Depending on the threat model and concept for chip-level implementation (Sec.~\ref{sec:threat_and_concept}), one may implement these dummy wires
	using RE-resilient interconnects in the
BEOL~\cite{patnaik17_Camo_BEOL_ICCAD},
	or with the help of additional MUXes and key bits to seemingly switch between real/dummy wires at the FEOL.
Similar protection is required for the assignment of the different input voltages
and control signals.

Finally, in addition to the 16 functions illustrated in Fig.~\ref{fig:GSHE_gates},
	we can readily extend our primitive to cloak latches and flip-flops, by applying the clock signal to
		the fixed ferromagnets' terminals.  Besides, the primitive can readily implement multi-input gates (i.e., $>$2 signal inputs)
as well.

\section{Threat Model and Concept for Secure Chip-Level Implementation}
\label{sec:threat_and_concept}

We assume the fab and the end-user to be untrusted; the ultimate
goal for any adversary is to understand the true functionality of a camouflaged/locked chip.
Our threat model represents a notable advancement over
prior work
related to camouflaging, where
the IP holder traditionally \emph{must} trust
the fab because of the device/circuit-level protection mechanism.

To hinder fab-based adversaries, we outline two equally promising options for secure implementation:
either (a) leverage split manufacturing~\cite{mccants11} or (b) provision
for a tamper-proof memory.
For option (a), the wires for the control inputs and the ferromagnet terminals shall remain protected from the untrusted FEOL fab. Hence, these wires have to be routed at least
		partially through the BEOL, which must be manufactured by a separate, trusted fab.
For option (b), the tamper-proof memory holds a secret key that defines (using some additional circuitry)
the correct assignment of control inputs and voltages for all devices.
The key must be loaded (by the IP holder) into the memory only after fabrication.

A malicious end-user can
obtain the design specifics of the chip through RE and side-channel attacks. She/he
can also 
use a working chip as an oracle for analytical attacks.
In the remainder of this paper, we focus on
malicious end-user.

\begin{table}[b]
\centering
\scriptsize
\caption{Characteristics of Synthesized
	Benchmarks
		(Italics: \emph{EPFL Suite}~\cite{EPFL15}; Bold: \emph{IBM Superblue Suite}~\cite{viswanathan2011ispd})
}
\label{tab:benchmarks}
\smallerspacecaption
\setlength{\tabcolsep}{0.7mm}
\begin{tabular}{c|c|c|c||c|c|c|c}
\hline
 \textbf{Benchmark
 } & \textbf{Inputs} & \textbf{Outputs} & \textbf{Gates
 } &
 \textbf{Benchmark
 } & \textbf{Inputs} & \textbf{Outputs} & \textbf{Gates 
 } \\ \hline \hline
\emph{aes\_{core}}  & 789 & 668 & 39,014 & \emph{log2} & 32 & 32 & 51,627 \\ \hline
b14 & 277 & 299 & 11,028 & \textbf{sb1} & 8,320 & 13,025 & 856,403 \\ \hline
b21 & 522 & 512 & 22,715 & \textbf{sb5} & 11,661 & 9,617 & 741,483 \\ \hline
c7552 & 207 & 108 & 4,045 & \textbf{sb10} & 10,454 & 23,663 & 1,117,846 \\ \hline
ex1010 & 10 & 10 & 5,066 & \textbf{sb12} & 1,936 & 4,629 & 1,523,108 \\ \hline
\emph{pci\_bridge32} & 3,520 & 3,528 & 35,992 & \textbf{sb18} & 3,921 & 7,465 & 659,511 \\ 
\hline
\end{tabular}
\end{table}

\section{Security Analysis}
\label{security}
\subsection{
	Study on Large-Scale IP Protection Against SAT Attacks}
\textbf{Setup:}
We model the proposed primitive and those of selected prior art~\cite{rajendran13_camouflage, parveen2017hybrid, alasad2017leveraging, zhang16, bi16_JETC, nirmala16, winograd2016hybrid, zhang2015giant} as outlined in~\cite{
	yasin15_IDT, massad15}.
Although the proposed primitive also supports locking,
	 here we contrast it only to camouflaging primitives;
logic locking and camouflaging are transformable notions without loss of generality~\cite{yasin15_IDT}.
Note that we also contrast to CMOS-centric techniques; this is meaningful as
any scheme hinges on the number and composition of their
cloaked functionalities~\cite{massad15,subramanyan15}, not their implementation (i.e., at least for
		analytical attacks).
For a fair evaluation, the same set of gates are protected---gates
are randomly selected once for each benchmark, memorized, and then reapplied across all techniques. 
We evaluate all techniques against powerful SAT attacks~\cite{subramanyan15, code_pramod, shen17},
run on
Intel Xeon server (2.3 GHz, 4 GB per task allowed).
The time-out (``t-o'') is set to 48 hours.

\textbf{Benchmarks:}
We conduct our experiments on traditional benchmarks suites (\emph{ISCAS-85}, \emph{MCNC}, and \emph{ITC-99}), on the
large-scale \emph{EPFL suite}~\cite{EPFL15}, and on the industrial \emph{IBM superblue} circuits~\cite{viswanathan2011ispd}
(Table \ref{tab:benchmarks}).
For the \emph{IBM superblue} circuits,
	we leverage~\cite{kahng14} to synthesize and generate the layouts for further analysis.
As for SAT attacks, we
pre-process the sequential circuits (\emph{IBM superblue}) as follows:
the inputs (and outputs) of all flip-flops become primary outputs (and inputs); thereafter, the flip-flops are removed.
(Doing so is essential to mimic access to scan chains for the SAT attacks~\cite{massad15}.)

\textbf{On provably secure versus large-scale schemes:}
Contrary to
\emph{provably secure} schemes such as~\cite{yasin16_SARLock, xie16_SAT,
	li16_camouflaging},
	one may find it difficult to engage in ``plain'' but large-scale camouflaging.
The key reason for this concern is that the solution space $C$---covering all possible functionalities of a camouflaged design and thereby defining the computational efforts for SAT
attacks---is hard to
quantify precisely~\cite{massad15,subramanyan15,li16_camouflaging}.
More specifically, $C$ depends primarily
on (i)~the number and the composition of functions cloaked by each primitive and
   (ii)~the number and selection of gates protected with a primitive.
Recall that prior art is limited in both (i) and (ii) by cost considerations.
In contrast, thanks to the innate polymorphism of the proposed primitive, we are unbound toward large-scale and even full-chip camouflaging.
Moreover, the primitive cloaks all 16 possible functionalities.
Intuitively, our scheme should thus impose maximal efforts for SAT attacks.
We believe that this renders our scheme competitive on par with
provably secure techniques, and
we substantiate this statement with a comprehensive study below.

\begin{table*}[tb]
\centering
\scriptsize
\setlength{\tabcolsep}{1mm}
\caption{Runtime for Our SAT Attacks (Using~\cite{subramanyan15,code_pramod}),
	in Seconds (Time-Out t-o is 172,800 Seconds, i.e., 48 Hours)
}\label{tab:satattackcomparison}
\smallerspacecaption
\begin{tabular}{*{14}{c|}c}
\hline
\multirow{3}{*}{\textbf{Benchmark}}
& \multicolumn{7}{|c|}{\textbf{10\% IP Protection}} & \multicolumn{7}{|c}{\textbf{20\% IP Protection}} \\
\cline{2-15}
& \textbf{\cite{rajendran13_camouflage}} & \textbf{\cite{nirmala16,winograd2016hybrid}}
& \textbf{\cite{bi16_JETC}}
	& \textbf{\cite[c]{alasad2017leveraging}, \cite{zhang16}} & \textbf{\cite{zhang2015giant}, \cite[a]{alasad2017leveraging}} &
	\textbf{\cite{parveen2017hybrid}} & \textbf{Our}
& \multirow{2}{*}{\textbf{\cite{rajendran13_camouflage}}} & \multirow{2}{*}{\textbf{\cite{nirmala16,winograd2016hybrid}}}
& \multirow{2}{*}{\textbf{\cite{bi16_JETC}$^\dag$}} & 
\multirow{2}{*}{\textbf{\cite[c]{alasad2017leveraging}, \cite{zhang16}}} & \multirow{2}{*}{\textbf{\cite{zhang2015giant}, \cite[a]{alasad2017leveraging}}} &
\multirow{2}{*}{\textbf{\cite{parveen2017hybrid}$^\ddag$}} & \multirow{2}{*}{\textbf{Our}}
\\
& (3)$^*$
& (6)$^*$
& (4)${^*}{^\dag}$
& (2)$^*$
& (4)$^*$
& (7+1)${^*}{^\ddag}$
& (16)$^*$
& 
& 
& 
& 
& 
& 
& 
\\ \hline
\hline
 
\emph{aes\_{core}}
& 610 & 4,710  & 890 & 132 & 536 & 6,229 & 25,890
& 4,319 & 41,844 & 11,306 & 407 & 9,432 & t-o & t-o
\\ \hline

b14
& 2,078 & 20,603 & 11,465 & 6,884  & 17,634 & 27,438 & 60,306
& 56,155 & t-o & 64,145 & 8,426 & t-o & t-o & t-o
\\ \hline

b21
& 7,813 & 162,324 & 45,465 & 3,977  & 24,035 & t-o & t-o
& t-o & t-o & t-o & t-o  & t-o & t-o & t-o
\\ \hline

c7552
& 37 & 210 & 74 & 12  & 66 & 371 & 2,289
& 169 & 14,575 & 1,153 & 110  & 1,327 & 172,548 & t-o
\\ \hline

ex1010
& 62 & 215 & 82 & 12  & 73 & 295 & 922
& 171 & 1,047 & 274 & 38  & 250 & 1,310 & 4,701
\\ \hline

\emph{log2}
& t-o & t-o & t-o & t-o  & t-o & t-o & t-o
& t-o & t-o & t-o & t-o  & t-o & t-o & t-o
\\ \hline

\emph{pci\_bridge32}
& 1,119 & t-o & 9,011 & 1,325  & 2,690 & t-o & t-o
& 54,577 & t-o & t-o & t-o  & t-o & t-o & t-o
\\
\hline

& \multicolumn{7}{|c|}{\textbf{30\% IP Protection}} &
	\multicolumn{7}{|c}{\textbf{40--100\% IP Protection$^\S$}}  \\
\hline
\hline
 
\emph{aes\_{core}}
& 17,148 & t-o & 31,601 & 2,020 & 26,498 & t-o & t-o & 
t-o & t-o & t-o & 8,206 & t-o & t-o & t-o
\\ \hline

b14
& 56,787 & t-o & t-o & 38,495 & t-o & t-o & t-o
& t-o & t-o & t-o & t-o & t-o & t-o & t-o
\\ \hline

b21
& t-o & t-o & t-o & t-o & t-o & t-o & t-o
& t-o & t-o & t-o & t-o & t-o & t-o & t-o
\\ \hline

c7552
& 1,786 & t-o & t-o & 766 & t-o & t-o & t-o
& t-o & t-o & t-o & 41,721 & t-o & t-o & t-o
\\ \hline

ex1010
& 448 & 4,357 & 938 & 87 & 719 & 11,736 & 24,727
& 1,703 & t-o & 129,290 & 169---7,073$^\S$  & 1,950 & t-o & t-o
\\ \hline

\emph{log2}
& t-o & t-o & t-o & t-o & t-o & t-o & t-o
& t-o & t-o & t-o & t-o & t-o & t-o & t-o
\\ \hline

\emph{pci\_bridge32}
& t-o & t-o & t-o & t-o & t-o & t-o & t-o
& t-o & t-o & t-o & t-o & t-o & t-o & t-o
\\ \hline

\end{tabular}
\\[1mm]
$^*$Number of cloaked functions; refer to Table~\ref{tab:devices} or the related publication for the actual sets of cloaked functions.
Prior art covering the same set is grouped into one column.
$^\dag$Here we refer to the camouflaging primitive, not the polymorphic gate reported on in Table~\ref{tab:devices}.
$^\ddag$Here
we also assume BUF to be available.
$^\S$The benchmark ex1010 can be resolved for 100\% IP protection, when the primitives of~\cite[c]{alasad2017leveraging}, \cite{zhang16} are used.
The related runtime range is for 40--100\% protection; all other runtimes are for
40\% protection.

\end{table*}

\textbf{Results:}
Table~\ref{tab:satattackcomparison} contrasts the resilience (against~\cite{subramanyan15,code_pramod}) of all considered schemes for large-scale application. For the same
	number of gates protected, we observe that the more functions a primitive can cloak, the more resilient it becomes in practice.
	More importantly, the runtimes required for decamouflaging (if possible at all), tend to scale exponentially with the percentage of gates being camouflaged.

	Our primitive induces by far the highest efforts across all
	benchmarks. Except for \emph{ex1010}, none of the benchmarks could be resolved within 48 hours once we protect 20\% or more of all gates.
	To confirm this superior resilience, we conducted further attacks running for 240 hours for full-chip protection using the proposed primitive---the designs could still not
	be resolved. Moreover, we also observe some
	computational failures;\footnote{E.g., ``\emph{internal error in 'lglib.c': more than 134,217,724 variables}''.}
this hints on another practical limitation w.r.t.\ scalability for SAT attacks, as one can reasonably expect~\cite{massad15}.

Besides the attacks of~\cite{subramanyan15,code_pramod}, we also leverage \emph{Double DIP}~\cite{shen17}.
The key advancement of this attack is that it rules out at least two incorrect keys in each iteration.
Conducting the very same set of experiments as in Table~\ref{tab:satattackcomparison}, we
observe that the runtimes are on average higher across all benchmarks.
For example, decamouflaging \emph{aes\_core} (for 10\% protection using our primitive) requires $\approx$7 hours using~\cite{subramanyan15}, but $\approx$15 hours using~\cite{shen17}.\footnote{Due to lack of space, we refrain from providing all detailed results on~\cite{shen17}.}
This finding suggests
	that large-scale camouflaging can be indeed on par with
	provably secure schemes.

Independent of our study, note that some prior art (e.g.,~\cite{winograd2016hybrid})
proposed cost-limited protection schemes. Here we have demonstrated that an overly
limited protection
cannot withstand powerful SAT attacks
(also recall Sec.~\ref{sec:background} for~\cite{winograd2016hybrid}).

\begin{figure}[tb]
\centering
\includegraphics[width=.95\textwidth]{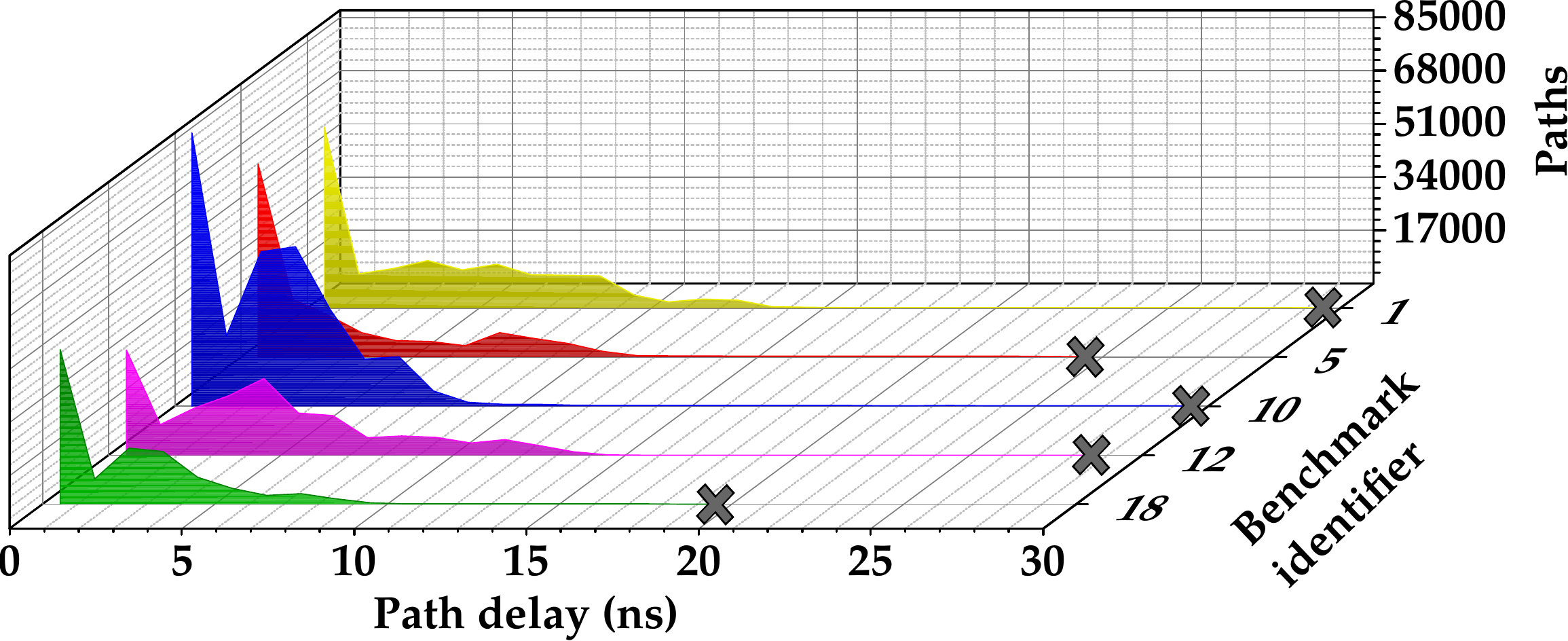}
\smallerspacecaption
\caption{Delay distributions of selected \emph{IBM superblue} circuits.
	The paths with the longest, critical delays are marked by crosses for clarity.
\label{fig:delay_superblue}
}
\end{figure}

Next, we outline the prospects for camouflaging of industrial circuits.
Recall that the delay of the GSHE switch is considerably higher when compared to CMOS (Sec.~\ref{device_model}).
Interestingly, large-scale circuits typically exhibit biased distributions of delay paths, with most paths having short delays but few paths having dominant, critical delays
(Fig.~\ref{fig:delay_superblue}).
In an experimental study on those \emph{IBM superblue} circuits, we replace CMOS gates in the non-critical paths with the GSHE-based primitive
such that no delay overheads can be expected.\footnote{We anticipate such hybrid designs to be practical, given the CMOS-compatible manufacturing of spin-based devices~\cite{Matsunaga2008
}. The main focus of this work, however, is
		hardware security, not circuit design.
	Hence, to mimic hybrid designs,
we replace the delay numbers of selected CMOS gates in non-critical timing paths with 
that of the GSHE switch, i.e., 1.55 ns.}
On an average, we can camouflage 5--15\% of all gates this way.
Conducting SAT attacks~\cite{subramanyan15,code_pramod} on those protected designs, we observe that they cannot be resolved within 240 hours; in fact, most runs incur similar
failures as discussed above. This indicates that the proposed primitive can help to strongly protect industrial circuits without excessive layout (PPA) overheads.

\subsection{On Stochastic Switching to Hinder SAT Attacks}

So far we leveraged the primitive in the context of classical, deterministic computation.
Note, however, that the underlying GSHE switch
supports
tunable
	probabilistic computation~\cite{rangarajan2017energy}.
Interestingly,
	the implications of probabilistic computation on hardware security are largely unexplored.

Recall the general principle of SAT attacks, i.e., 
to carefully apply input patterns on a working chip
and to observe the output patterns, throughout multiple sampling iterations, until the correct assignment for all key bits can be derived (by ruling out incorrect keys via disagreement).
Now consider a scenario where the GSHE switch (or any probabilistic device, for that matter)
is tuned for 95\% accuracy.
This implies that 5\% of the patterns observed by the SAT attack are incorrect.

We believe that most if not all proposed SAT attacks will fail in such scenarios.\footnote{The most promising contender here is arguably \emph{AppSAT}~\cite{shamsi17}, which is based
	on the probably-approximately-correct (PAC) paradigm.
		The attack as outlined in~\cite{shamsi17}, however, requires a consistent solution space regarding the input-output queries---probabilistic computation violates this
		assumption. The attack was not available to us for an experimental study at this time.}
That is because they have not been tailored to account for incorrect output patterns.
Even if they were, distinguishing incorrect patterns from correct ones is difficult when only given a ``probabilistic black-box oracle.''
Naturally, one might want to leverage machine learning (ML) toward this end.
We argue, however, it remains to be seen whether ML-based attacks will be sufficiently robust and capable.
Here we like to point out that
(i)~the GSHE switch experiences
thermally induced stochasticity~\cite{rangarajan2017energy}, (ii)~the error rate for any switch can be tuned individually, and
(iii)~those individual distributions superpose with each other while they propagate throughout the entire design, resulting in stochastically correlated behavior at the primary
outputs.

\subsection{Preventing Reverse Engineering and Side-Channel Attacks}

\textbf{Layout identification and read-out attacks:}
   Recall that the
   layout of the proposed primitive is uniform (Sec.~\ref{device_model}), hence indistinguishable for optical-imaging-based RE.
		A more sophisticated 
attacker might, however, leverage electron microscopy (EM) for identification and read-out attacks~\cite{courbon16}.
While such
attacks are yet to be demonstrated on switching devices at runtime,
we believe that
the proposed primitive can prevent them.
First, the dimensions of the GSHE switch are significantly smaller than CMOS devices, which
is a challenge regarding the spatial resolution for EM-based analysis~\cite{courbon16}.
Second, the primitive is truly polymorphic, i.e., its functionality can be switched at runtime; see also next.

\textbf{Polymorphism at the chip-level:}
Given truly polymorphic gates and some circuitry to judiciously switch the functionalities of gates,
		we can implement \emph{runtime polymorphism} at the chip-level.
Then, internal functionalities are not static (possibly even for static input patterns),
whereupon an RE-centric attacker is bound to misinterpret parts of the layout---it is virtually impossible to resolve all dynamic features on full-chip scale at once.\footnote{In~\cite{courbon16}, e.g., it took 50 ns to read-out one pixel of one memory cell, which is well above
the 1.55 ns speed of the GSHE device.}
Independent of RE threats, runtime polymorphism at the chip-level can also enable dynamic protection, e.g., as recently proposed by Koteshwara \emph{et al.}~\cite{koteshwara17}.
Their idea is to
alter the key dynamically, thereby rendering runtime-intensive attacks incapable (SAT attacks in particular).

\textbf{Photonic side-channel attacks:} 
While CMOS devices emit photons during operation, making them vulnerable to powerful attacks such as~\cite{Schlösser2012},
the GSHE switch itself
does not emit any photons.
The fundamentally different switching principle
hence makes the proposed primitive inherently resilient to read-out attacks based on photons.
Still,
	we caution that an assessment against such attacks
	shall be performed in future.

\textbf{Magnetic and temperature attacks:} 
Ghosh~\emph{et al.}~\cite{ghosh2016spintronics}
outlined attacks on spintronic (memory) devices using magnetic 
fields and temperature curves.
The design of the GSHE switch shall ensure a robust coupling between the W and R nanomagnets~\cite{rangarajan2017energy}.
This would naturally be disturbed by any external magnetic fields.
Hence, an attacker leveraging a magnetic probe may induce stuck-at-faults which are, however,
hardly controllable due to multiple factors (very small size of switches, accordingly large magnetic fields required for the probe, state of W and R magnetizations, the orientation of the fixed magnets, voltage polarities on the fixed magnets).
This implies that sensitization attacks such as~\cite{rajendran13_camouflage} will be difficult, if practical at all.
Regarding temperature-driven attacks, note that the retention time of the switch will be impacted. The resulting disturbances, however, are likely stochastic due to the inherent thermal noise in the nanomagnets.

\section{Conclusion}
\label{conclusion}
We explore the
security aspects of
	the GSHE switch: a versatile spin-based polymorphic device
which can support both camouflaging and logic locking.
Through a comprehensive study using SAT attacks, we show the strong resilience of our deterministic primitive as compared to
prior art.
We further discuss the resilience of our primitive against various classes of side-channel attacks.
Finally, we lay the foundations for promising security concepts:
	truly polymorphic behavior at runtime, and
	stochastic behavior to thwart analytical attacks.

\section*{Acknowledgements}
	This work was carried out in part on the High Performance Computing resources at New York University Abu Dhabi.

\bibliography{main}
\bibliographystyle{IEEEtran} 

\ifCLASSOPTIONcaptionsoff
  \newpage
\fi

\end{document}